\documentclass[12pt]{iopart}

\usepackage{graphics}

\def\PRA{{\it Phys.~Rev.~A} }

\def\JPB{{\it J.~Phys.~B} }
\def\PRL{{\it Phys.~Rev.~Lett.} }

\newcommand{\myscalebox}[1]{\scalebox{0.5}[0.5]{#1}}
\newcommand{\myscaleboxa}[1]{\scalebox{0.6}[0.5]{#1}}
\newcommand{\myscaleboxb}[1]{\scalebox{0.6}[0.4]{#1}}
\newcommand{\myscaleboxc}[1]{\scalebox{0.6}[0.6]{#1}}

\begin{document}

\title[Retrieval of interatomic separations of molecules from high-order harmonic spectra]
{Retrieval of interatomic separations of molecules from
laser-induced high-order harmonic spectra}

\author{Van-Hoang Le$^1$, Ngoc-Ty Nguyen$^1$, C. Jin$^2$, Anh-Thu Le$^2$ and C.~D. Lin$^2$}

\address{$^1$ Department of Physics, University of Pedagogy, 280 An Duong Vuong,
Ward 5, Ho Chi Minh City, Vietnam}

\address{$^2$ J. R. Macdonald Laboratory, Department of Physics, 
Kansas State University, Manhattan, Kansas 66506, USA}

%\ead{atle@phys.ksu.edu}

\begin{abstract}
   We illustrate an iterative method for retrieving the internuclear separations
   of  N$_2$, O$_2$  and CO$_2$ molecules using the high-order harmonics generated  from
   these molecules by intense infrared laser pulses. We show that accurate results
   can be retrieved with a small set of harmonics and with one or few alignment
   angles of the molecules.  For linear molecules the internuclear
   separations can also be retrieved from harmonics generated using isotropically
   distributed molecules. By extracting the transition dipole moment from the
   high-order harmonic spectra, we further demonstrated that it is preferable
   to retrieve the interatomic separation iteratively by fitting   the extracted
   dipole moment. Our results show that time-resolved chemical imaging of molecules
   using infrared laser pulses with femtosecond temporal resolutions is possible.

\end{abstract}

%Uncomment for PACS numbers title message
% Keywords required only for MST, PB, PMB, PM, JOA, JOB?
%\vspace{2pc}
%\noindent{\it Keywords}: Article preparation, IOP journals
% Uncomment for Submitted to journal title message
\submitto{\JPB}
% Comment out if separate title page not required
\maketitle

\section{Introduction}
In recent years, it has been the dream of physical and chemical
scientists to understand the intermediate steps of chemical
reactions or biological transformations \cite{zewail00,zewail01}.
For temporal resolutions of the order of subpicoseconds the
conventional x-ray and electron diffraction methods are not
suitable for such time-dependent imaging studies. Today infrared
lasers of durations of tens to sub-ten femtoseconds are widely
available, thus it is natural to ask whether infrared laser pulses
can be used for dynamic chemical imaging of molecules. (For
earlier attempts and results, see the recent review
\cite{leinreview}). When an atom or molecule is exposed to an
intense laser pulse, the electrons that were released earlier may
return to recombine with the target ion with the emission of
high-order harmonics. Because   recombination occurs when the
returning electrons are near the target ion,  high-order harmonic
generation (HHG) spectra thus contain information on the structure
of the target. In a recent paper, Itatani {\it et al}
\cite{itatani} reported that they have successfully reconstructed
the highest occupied molecular orbital (HOMO) of N$_2$ molecules
from the measured HHG spectra using the tomographic procedure.
This widely cited paper has generated a lot of interest since it
points out the opportunity for time-resolved imaging of transient
molecules using infrared lasers, with temporal resolutions of tens
to sub-ten femtoseconds, depending on the duration of the probe
pulse.

The tomographic procedure reported in Itatani {\it et al}
\cite{itatani} employed a number of assumptions. Before the method
can be generally implemented, the underlying assumptions should be
carefully analyzed. In a recent theoretical paper \cite{hoang}, it
has been shown that the tomographic procedure relies on the crude
approximation that the returning electrons be treated by plane
waves. For photo-recombination processes (or its time inverse, the
photoionization process) it is well-known that plane waves are
very poor approximations for describing a continuum electron in
the target ion field for energies in the energy range of tens to
hundreds eV's. In spite of this, the basic idea laid out in
Itatani {\it et al} \cite{itatani} of retrieving the structural
information from the HHG remains very attractive. In
Ref.~\cite{hoang}, it was argued that it is not essential to
extract the HOMO in order to ``know'' the structure of molecules.
For a molecule under transformation, often the bond lengths and
bond angles would evolve in time.  If the interatomic positions of
all the atoms in a transient molecule can be retrieved at each
given time, the goal of dynamic chemical imaging is mostly met.
For this purpose, in Le {\it et al} \cite{hoang} it was proposed
to retrieve the interatomic distances from the HHG spectra
directly using an iterative procedure.

The main purpose of this paper is to illustrate in practice how
the iterative method works. The paper is organized as follows. In
Section 2, we will briefly describe the theoretical basis of the
tomographic procedure and the iterative method. The main results
will be presented in Section 3. We first show an example of
extracting structural information of CO$_2$ using the tomographic
procedure. We then present results of the fitting procedure, which
is one of the main ingredients of the iterative method, on example
of N$_2$ and CO$_2$. By retrieving the interatomic separations
only, we will show that they can be extracted from HHG experiments
even when the molecules are isotropically distributed. We will
also show that the fitting procedure can be applied directly to
the dipole moments instead of the HHG yields. This offers an
opportunity for simple extension beyond the plane-wave and
single-active electron approximations. The last section summarizes
our conclusions and perspective of further studies. Atomic units are
used throughout unless otherwise indicated.

\section{Theoretical Methods}

\subsection{The tomographic procedure for retrieving HOMOs}
Based on the three-step model, in Itatani {\it et al}
\cite{itatani},  the HHG yield from a molecule in a laser field is
written approximately as

\begin{equation}
S(\omega,\theta) \sim N (\theta)\omega^4|a[k(\omega)]{\bi d}(\omega,\theta)|^2
\end{equation}
Here ${\bi d}(\omega,\theta)$ is the transition dipole between the
valence molecular orbital (or HOMO) and the continuum state;
$a[k(\omega)]$ is the amplitude of the continuum wave of the
returning electrons; $N(\theta)$ is the angular dependence of the
tunneling ionization rate for molecules aligned, where $\theta$ is
the angle between the molecular axis and the laser polarization
direction. Using the fact that in the tunneling regime, the
returning electron wave packet depends weakly on the target, one
can eliminate $a[k(\omega)]$  by measuring the HHG from a
reference atom with similar ionization potential. More precisely,
the transition dipole of the molecule can be extracted from
experimentally measured HHG spectra by

\begin{equation}
|{\bi d}(\omega,\theta)| \sim N(\theta)^{-1/2}|{\bi
d}_{ref}(k)|\sqrt{S(k,\theta)/S_{ref}(k)}
\end{equation}
with ${\bi d}_{ref}(k)$ and $S_{ref}(k)$ being the transition
dipole and HHG yield for the reference atoms, respectively.

   Thus the starting point of Itatani {\it et al} \cite{itatani} is to extract the dipole
   matrix element of a molecule by measuring the HHG spectra for
   molecules aligned at different angles $\theta$ with respect to the HHG
   spectra generated from the reference atom. Assuming that the dipole
   matrix element of the reference atom is known, and that the angular
   dependence of tunneling ionization rate $N(\theta)$ can be obtained from the
   MO-ADK theory \cite{moadk}, Eq.~(2) allows the determination of the dipole
   matrix elements if the dipole matrix element is taken to be real
   or purely imaginary (but not necessarily positive definite).

  To extract the HOMO orbital using the tomographic procedure, many
  additional assumptions have to be made, the most critical one being
  that the dipole matrix element ${\bi d}(\omega,\theta)$ be evaluated
  between the HOMO ground
  state wave function and continuum wavefunctions,  where the latter
  are approximated by plane waves. Under this assumption the dipole
  matrix elements are just the weighted Fourier transform of the HOMO
  wavefunction. By performing the inverse Fourier transform the HOMO
  wavefunction are thus extracted.  From the operational viewpoints,
  additional assumptions had been made in Itatani {\it et al} \cite{itatani}.
  The Eq.~(1) above was written for HHG generated by a single target molecule.
  In Itatani et al, the HHG spectra were obtained from experiment such
  that propagation in the medium has to be accounted for. They assumed
  perfect phase matching, such that $N(\theta)$ in Eq.~(1) is replaced by
  $N(\theta)^2$. Eq. (1) was also written for molecules fixed in space.
  They used a
  weak aligning laser pulse to partially align the molecules. For
  photo-recombination, the electron momentum and photon energy is
  related by $k=\sqrt{2(\omega-I_p)}$  where $I_p$ is the ionization
  energy of the target. In Itatani {\it et al} \cite{itatani},
  a different dispersion with $k=\sqrt{2\omega}$ was used in order
  to extract the HOMO orbital more accurately.  Furthermore, to use
  the tomographic procedure,  both polarizations of the emitted HHG's
  should be measured. In Itatani {\it et al} \cite{itatani}, the HHG
  yields with the perpendicular polarization were not measured.

\subsection{The iterative procedure for retrieving interatomic
separations}

   The tomographic procedure, as summarized above, in addition to
   the many assumptions required above, also needs the deduced dipole
   moment over the whole energy range in order to perform the inverse
   Fourier transform. In high-order harmonic generation, the harmonic
   yield drops precipitously beyond the cutoff and thus the dipole
   moment in the high photon energy is not available.

   In Le {\it et al} \cite{hoang}, it was argued that it is not essential to extract
   the HOMO orbital in order to infer the structure of the molecule.
   The perceived applications of using infrared lasers for probing
   the structure of molecules owes to the femtoseconds temporal
   resolutions offered by these laser pulses. To follow the time-evolution
   of a molecule under transformation, it is of foremost importance to
   specify how the positions of its atomic centers change in time.
   Thus we perceive that our main task is to extract the internuclear
   distances from the HHG photons generated by applying a probe laser.

   Determination of molecular structure from the measured HHG is a kind
   of inverse scattering problem. Assume that we know how to calculate
   the HHG's exactly if the molecular structure is known, the retrieval
   problem is to find a procedure where the structure can be determined
   from the measured HHG's.  Operationally, we will assume that the
   initial configuration of the molecule in the ground state
is known from the conventional imaging method. Under chemical
transformation, our goal is to locate the positions of the atoms
as they change in time, by measuring their HHG spectra at
different time delays, following the initiation of the reaction.
Assuming that such HHG data are available experimentally, the task
of dynamic chemical imaging is to extract the intermediate
positions of all atoms during the transformation and, in
particular, to identify the important transition states of the
reaction. This will be done by the iterative method.

In the iterative method, we will first make a guess as to the new
positions of all atoms in the molecule. A good guess is to follow
the reaction coordinates along the path where the potential
surfaces have the local minimum. Existing quantum chemistry codes
(see, for example, \cite{gaussian, gamess}) provide good guidance
as a starting point. For each initial guess the HHG will be
calculated. The resulting macroscopic HHG is then compared to the
experimental data to find atomic configurations that best fit the
experimental data.

Since experimental and accurate theoretical HHG spectra for
molecules are not readily available, here we generated our
``experimental'' data using the strong-field approximation (SFA)
(or the Lewenstein model) \cite{lewenstein}. To illustrate the
method, we will limit ourselves to simple linear molecules only.
In this case, the iterative method can be implemented in a
straightforward manner, as the only parameter involved is the
internuclear distance. Consider a diatomic molecule like N$_2$,
and a linear triatomic molecule CO$_2$. For a given internuclear
separation $R_0$ (for CO$_2$, it is the distance between oxygen
and carbon), we use the SFA to calculate the HHG spectra. We then
take the calculated results and introduce random errors of the
data on each harmonic order of up to $50\%$, to simulate random
``experimental'' uncertainty.  Assuming that the internuclear
separation is not known, we generate new HHG data
$S^{th}(\theta,\omega,R)$ using a range of internuclear
separations. By minimizing the calculated variance, we show that
the correct internuclear separation $R_0$ can be retrieved.
Specifically, for each molecular alignment, we calculate the
variance

\begin{equation}
\sigma(\theta,R)=\sum_{\omega_{min}}^{\omega_{max}}
\left\{\log\left[S^{exp}(\theta,\omega,R_0)\right]
-\log\left[S^{th}(\theta,\omega,R)\right]\right\}^2
\end{equation}
for various $R$ near some initial guess value to find the
minimum of $\sigma(\theta,R)$. Here the harmonic spectra are
calculated using the SFA, extended for molecular systems \cite{zhou,
atle}, with the HOMO for each $R$ obtained from the {\it Gamess} code
\cite{gamess}. For each odd harmonic order $(2n+1)$, the yield is
obtained by summing the yield for the frequency range from order
$2n$ to $(2n+2)$. In the equation above, the ``experimental data''
$S^{exp}(\theta,\omega,R_0)$
already include the random uncertainty, and the summation is over
a frequency range chosen, typically from H11 up to the harmonic
cutoff.

Since only one parameter is being retrieved, the data set needed
is small.  It will be shown that $R_0$ can be determined
accurately using only a small set of harmonics, and that there is
no need to measure HHG for many different alignment angles of the
molecules. Furthermore, the fitting procedure also allows us to
retrieve the internuclear separation accurately in case of
isotropically distributed molecules.

 Clearly the fitting procedure can be applied directly to the
 transition dipole  extracted using Eq.~(2) as well. Assume that the
 ``experimental'' transition dipole is available from Eq.~(2). In the next step,
 which deviates from the tomographic procedure, one calculates the
 transition dipole for a guess value of $R$ and compare with the
 dipole extracted from experiments. The process continues until
 one gets the best fit. For diatomic molecules, the method is very
 simple as one can easily scan over a range of internuclear distance
 (the only parameter). Similar to the fitting to HHG spectra [Eq.~(3)],
 one can look for the minimum in the variance of the transition dipoles

\begin{equation}
\sigma(\theta,R)=\sum_{\omega_{min}}^{\omega_{max}}
\left[\left|Cd^{exp}(\theta,\omega,R_0)\right|
-\left|d^{th}(\theta,\omega,R)\right|\right]^2
\end{equation}
Since transition dipole typically behaves linearly as a function
of energy (or HHG order), there is no need to use the logarithmic
function as in case of fitting to the HHG spectra [Eq.~(3)]. Note
that we have also introduced an additional fitting coefficient $C$
to account for the overall normalization factor in the
``experimental'' dipole, which is not easily fixed [see Eq.~(2)].
Since the variance depends quadratically on $C$, the condition for
minimum $\partial\sigma/\partial C=0$ gives

\begin{equation}
C=\frac{\sum_{\omega_{min}}^{\omega_{max}}
|d^{exp}(\theta,\omega,R_0)d^{th}(\theta,\omega,R)|}
{\sum_{\omega_{min}}^{\omega_{max}}|d^{exp}(\theta,\omega,R_0)|^2}
\end{equation}
With $C$ fixed by the above equation, one needs only minimize the variance
with respect to $R$.

\section{Results and Discussion}

\subsection{Tomographic method for CO$_2$}

Here we demonstrate the tomographic procedure in an example of
CO$_2$. This is to complement the examples of
 N$_2$ and O$_2$ presented in \cite{hoang}, also because this system has been
 examined in a few experiments recently \cite{levesque,NRC}.
For the reference atom, we use Kr, with ionization potential of
$13.99$ eV which is very close to the one of CO$_2$ ($13.78$ eV).
We use SFA \cite{lewenstein,zhou,atle} to generate the
``experimental'' HHG data   used in Eqs.~(1) and (2). Without loss
of generality, we assume that the molecule is aligned along the
$x$ axis in a laser field, linearly polarized on the $x$-$y$ plane
with an angle $\theta$ with respect to the molecular axis. A 30 fs
(FWHM) laser pulse with  peak intensity of $2\times10^{14}$
W/cm$^2$ and wavelength of $1200$ nm is used.

\begin{figure}
\centering
\mbox{\rotatebox{0}{\myscaleboxa{
\includegraphics{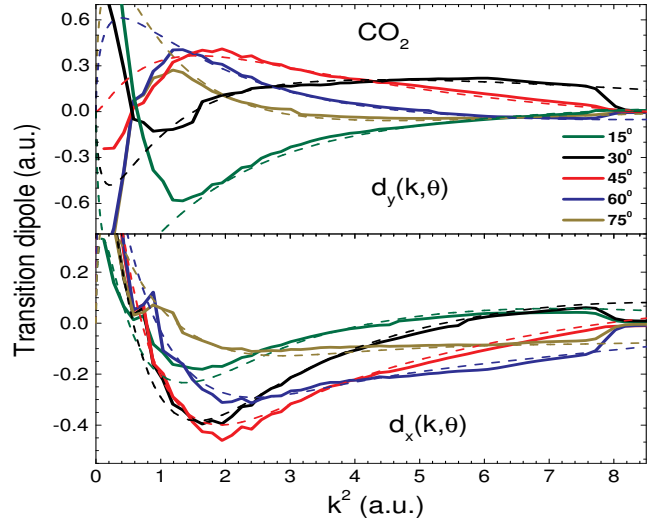}}}}
\caption{The extracted transition dipoles of CO$_2$ (solid lines)
in comparison with the theoretical ones (dash lines) for few
different alignments, shown in the labels. The upper and lower
panels are for the $y$ and $x$ components, respectively. A 30 fs
(FWHM) laser pulse with peak intensity of $2\times 10^{14}$
W/cm$^2$ and wavelength of $1200$ nm is used.} \label{fig1}
\end{figure}

In Fig.~1 we compare
the dipole moments extracted by using Eq.~(2) with the theoretical
data for a few alignment angles shown on the label. Clearly, one
can see that the extracted data compare quite well with the
theoretical ones for a broad range of $k^2\in [1-8]$ a.u., or from
harmonic orders H27 to H109.  Note that we choose the long
wavelength of $1200$ nm instead of $800$ nm in order to have a
broader useful range of harmonics for the tomographic procedure,
as has been suggested in \cite{hoang}. With these extracted
dipoles, we then use the Fourier slice theorem to calculate the
HOMO wavefunction. The contour plot of the retrieved HOMO for the
case of equilibrium CO$_2$ shown in Fig.~2(e) has a clear $\pi_g$
symmetry, which compares quite well with the theoretical contour
plot shown in Fig.~2(b). The distance between the oxygen and
carbon centers, estimated as half the distance between the peaks
along $x$-axis, is $R^*=2.12$ au, is also in reasonable good
agreement with the input $R_0=2.2$ au. Similarly, good agreements
are found for the case of $R_0=1.61$ and $2.79$ au, shown in the
upper and lower panels of Fig.~2, respectively. The retrieved
distances are $R^*=1.71$ and $2.63$ au, respectively.

\begin{figure}
\centering
\mbox{\rotatebox{0}{\myscalebox{
\includegraphics{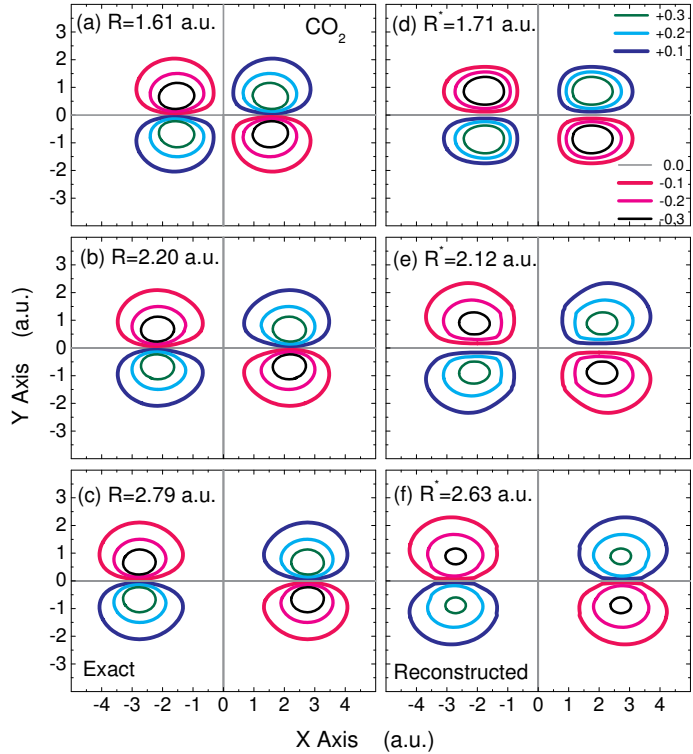}}}}
\caption{Contour plot of the retrieved HOMO wavefunctions of CO$_2$
[(d),(e),(f)], as compared to the exact ones [(a),(b),(c)] with the input
internuclear distances (between oxygen and carbon centers) $R_0=1.61$,
$2.2$, and $2.79$ a.u., respectively.} \label{fig2}
\end{figure}

\begin{figure}
\centering
\mbox{\rotatebox{0}{\myscaleboxc{
\includegraphics{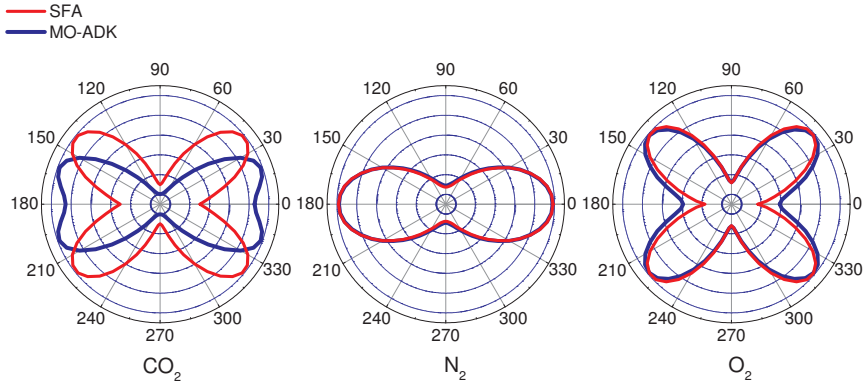}}}}
\caption{Polar plot of the ionization rates for CO$_2$, N$_2$ and O$_2$
from MO-ADK and SFA calculations, as functions of alignment angle between
molecular axis and laser polarization direction. Laser pulse with
peak intensity of $2\times10^{14}$ W/cm$^2$, mean wavelength of
$800$ nm and duration (FWHM) of 30 fs was used in calculations.}
\label{fig3}
\end{figure}

We found that in case of CO$_2$, the extracted dipoles agree
better with the theoretical ones if the strong-field approximation
is used to calculate ionization rate instead of the simple MO-ADK
theory. It is interesting to note that recent measurements by the
NRC group \cite{NRC} also showed the alignment dependence of the
ionization rate for CO$_2$ deviated significantly from the
prediction by the MO-ADK theory. On the other hand, the MO-ADK
predictions for N$_2$ and O$_2$  appear to be in very good
agreement with the SFA results and with experiments
\cite{NRC,litvinyuk,cockeN2}. In Fig.~3, we compare the CO$_2$
ionization rates from MO-ADK theory (blue line) and SFA (red
line), plotted as function of  the alignment angle. A laser with
peak intensity of $2\times10^{14}$ W/cm$^2$, mean wavelength of
$800$ nm and duration (FWHM) of $30$ fs is used. The MO-ADK rate
peaks near $30^{\circ}$, which is in agreement with results
  deduced from measured double
ionization of CO$_2$ \cite{cockeCO2}. For MO-ADK calculations, we
use the values of the coefficients $C_l$ as suggested in
\cite{atle07}.  The SFA rate for CO$_2$ is quite similar in shape
to that from O$_2$, but with the peak shifted to about
$37^{\circ}$. This result appears to be in a better agreement with
the NRC measurements, which show very narrow peak near
$45^{\circ}$. The nature of the discrepancies between the two
theories and with experiments are not clear so far. We note,
however, that the good retrieval results by the tomographic
procedure with the SFA theory for the ionization rates is likely
due to the fact that the SFA model was also used to generate the
HHG data. For completeness, we show the comparison between SFA and
MO-ADK theories in case of N$_2$ and O$_2$, in Fig.~3(b) and (c),
respectively. Clearly, the two theories agree quite well here.

\subsection{Fitting procedure for extracting internuclear distances
in case of fixed alignments}

Now we discuss the results from the fitting procedure, described
in Sec.~2.2. The success of the fitting procedure for extracting
internuclear distances $R_0$ requires that HHG spectra be sensitive
to this parameter.  In Fig.~4(a) we show the calculated HHG from
single N$_2$ molecules by a 30 fs laser pulse of wavelength of 800
nm and peak intensity of $2\times10^{14}$ W/cm$^2$. The molecule is
assumed to be aligned at $\theta=0^{\circ}$ with respect to the
laser's polarization axis. The known
equilibrium distance for N$_2$ is $2.09$ a.u. We also show the HHG
spectra with inputs $R_0=1.84$ and 2.27 a.u., amounting to about
$10\%$ decrease or increase from the equilibrium distance,
respectively. In Fig.~4(a) one can see that the HHG spectra for
the three distances are quite distinguishable. The noticeable
minimum for each $R_0$ moves to lower HHG order as $R_0$ is
increased. In Fig.~4(b) we show that an introduction of $30\%$ or
$60\%$ ``experimental'' errors to the data from $R_0=2.09$ a.u.
does not show noticeable difference as compared to the actually
calculated spectra. This comparison sheds light that the retrieval
of $R_0$ should be relatively straightforward.

\begin{figure}
\centering
\mbox{\rotatebox{0}{\myscaleboxa{
\includegraphics{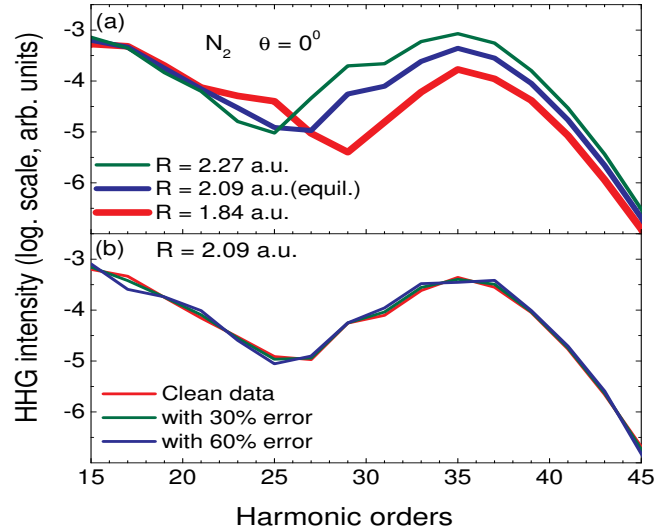}}}}
\caption{(a) HHG from N$_2$ with different internuclear distances
$R_0=2.27$ (green line), $2.09$ (blue line), and $1.84$ (red line)
calculated with fixed alignment angle $\theta=0^{\circ}$. (b) HHG
from N$_2$ in equilibrium internuclear distance ($R_0=2.09$ au)
with random ``experimental'' errors of $30\%$ (green line) and $60\%$
(blue line) at each harmonic.} \label{fig4}
\end{figure}

\begin{figure}
\centering
\mbox{\rotatebox{0}{\myscaleboxa{
\includegraphics{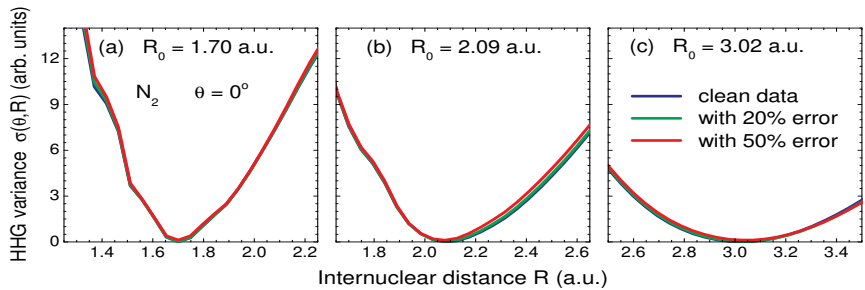}}}}
\caption{Variance of the HHG from N$_2$ for $R$ chosen near the
input $R_0$ for the  three different inputs $R_0 =1.70$, $2.09$
(equilibrium) and $3.02$ a.u.} \label{fig5}
\end{figure}

\begin{figure}
\centering
\mbox{\rotatebox{0}{\myscaleboxa{
\includegraphics{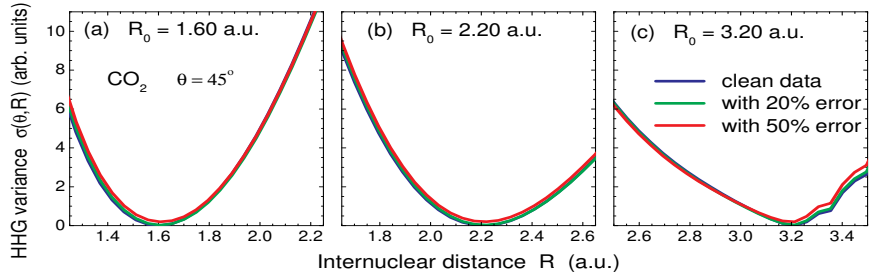}}}}
\caption{Same as Fig.~5, but for CO$_2$ with inputs $R_0 =1.6$,
$2.2$ (equilibrium) and $3.2$ a.u.} \label{fig6}
\end{figure}

In Fig.~5, we show the variance as a function of internuclear
distance, calculated by using Eq.~(3), for each test case of N$_2$
with input $R_0=1.7, $ $2.09$, and $3.0$ au. Clearly the minimum
in each case occurs near the input $R_0$, meaning that the value
of the input internuclear separation can indeed be extracted with
this simple fitting procedure. In the calculations shown in
Fig.~5, we use the range of HHG in the plateau region, from
H15-H45 for the alignment angle $\theta=0^{\circ}$. Similar
results for CO$_2$ are shown in Fig.~6 using the same laser
parameters, but for $\theta=45^{\circ}$.  Again, the position of
the minimum in each case matches closely to the input internuclear
distance (between oxygen and carbon centers) of $R_0=1.6$, $2.2$,
and $3.2$ a.u. Note that, in these calculations, only harmonics with
parallel polarization are used. For the tomographic procedure,
strictly speaking, one needs to know both components. We have
tested the results by using different alignment angles and
different ranges of harmonics and the retrieved internuclear
distances are all very close to the input ones. Note that the
accuracy of the retrieved internuclear distance can also be
checked by using lasers of different wavelength or intensity.

\subsection {Fitting in case of isotropic molecular distributions}

The above fitting procedure has been applied to molecules fixed in
space. At finite temperature, molecules can only be partially
aligned or oriented. The above procedure can be generalized to
these partially aligned molecules.  Since only the internuclear
distances are extracted, the fitting procedure can also be applied
to molecules that are isotropically distributed. To illustrate the
method, in this paper we assume that
experimental HHG spectra from such isotropically distributed
molecules are produced fully in phase so that the HHG amplitudes
are added coherently.

\begin{figure}
\centering
\mbox{\rotatebox{0}{\myscaleboxb{
\includegraphics{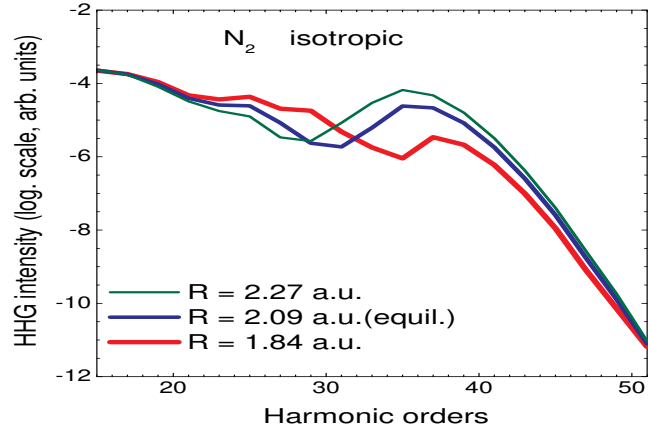}}}}
\caption{Same as Fig.~4(a), but for isotropically distributed
molecular N$_2$. Such data can be used to retrieve $R_0$ using the
fitting procedure.} \label{fig7}
\end{figure}

\begin{figure}
\centering
\mbox{\rotatebox{0}{\myscaleboxa{
\includegraphics{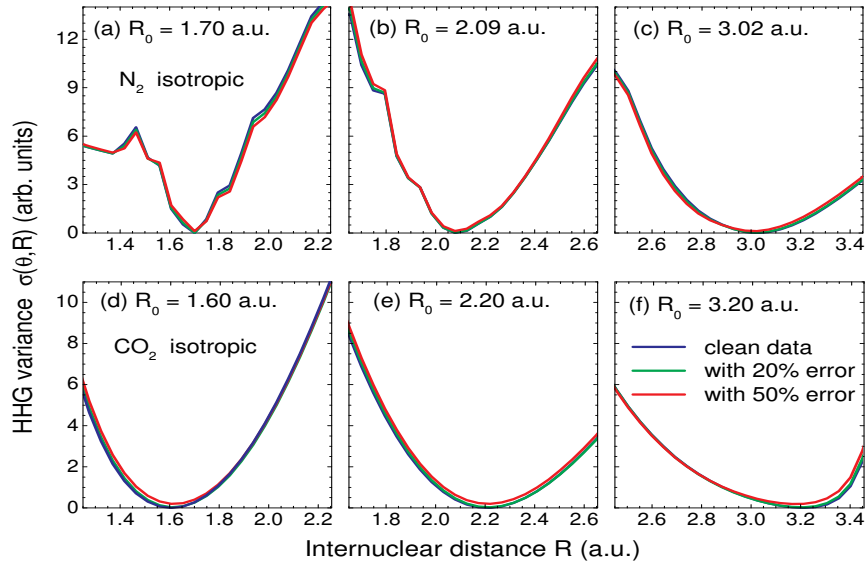}}}}
\caption{(a),(b), and (c): Variance of the HHG from N$_2$ for $R$
near the input $R_0=1.70$, $2.09$, and $3.02$ a.u., respectively,
for the case of isotropic distribution. (d), (e), and (f): Same as
(a), but for CO$_2$ near $R_0=1.60$, $2.20$, and $3.20$ a.u.,
respectively.} \label{fig8}
\end{figure}

In Fig.~7, we show the calculated HHG spectra for such
isotropically distributed N$_2$ molecules at $R_0=1.84$, $2.09$,
and $2.27$ a.u. Note that the pronounced minima in the spectra can
still be seen clearly, although their positions are moved to
somewhat higher harmonics, compared to the case of
$\theta=0^{\circ}$, shown in Fig.~4(a). This is not surprising
since the HHG yield for small $\theta$ is dominant. Using the
same fitting procedure [Eq.~(3)], we confirmed that the $R_0$ can
be extracted with similar accuracy. Typical results are
presented in Fig.~8(a) and (b) for N$_2$ and CO$_2$, respectively,
for few inputs $R$. Thus for simple systems, the
internuclear distance can be extracted from the HHG yields even if
the molecules are not aligned.

\subsection{Fitting to transition dipoles}

The above results establish the basic framework for the fitting
procedure. The method is formulated in a general form and could
easily be changed to tailor with modifications and extensions. We now
discuss the results from the fitting to the transition dipoles. In
Fig.~9 we show variance $\sigma(\theta,R)$  for transition dipole of O$_2$
with the input internuclear distance of $R_0=2.29$ a.u. and
alignment angle of $45^{\circ}$. The fitting equation (4) is used,
with the harmonics generated by the laser with peak intensity of
$2\times10^{14}$ W/cm$^2$, wavelength of $800$ nm and duration of
$30$ fs. The summation in Eq.~(4) was carried out for the range H19-H43.
That corresponds to the fitting range $k^2\in [1.28-4.0]$ a.u.
The ``experimental'' transition dipoles were calculated
using Eq.~(2), using Xe as the reference atom. For each guess
of the internuclear distance, the
theoretical dipoles are calculated within the plane-wave
approximation for the continuum electron, in order to be
consistent with the SFA model for the HHG used here. As can be
seen from the figure, the minimum position indeed agrees well with
the input $R_0$. The data are shown for the fitting to the
$d_x$ component of the dipole (in the molecular frame). The fitting
to $d_y$ component leads to similar result. In fact, the minimum positions
are very close to the input $R_0=2.29$ a.u. for all other alignment angles
we tried.

\begin{figure}
\centering
\mbox{\rotatebox{0}{\myscaleboxb {
\includegraphics{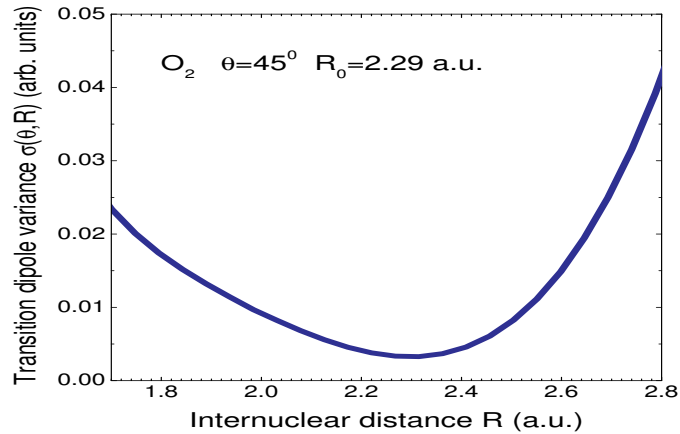}}}}
\caption{Variance of the transition dipole for O$_2$ for $R$ near
the input $R_0=2.29$ au.} \label{fig9}
\end{figure}

 Note that dipole moments were used directly in
extracting the structural information in this approach, similar to
the method used in the tomographic procedure. Once again, since
only the internuclear distance is retrieved, the method is much
easier to implement experimentally in time-resolved measurements.
In real situation, in order to fit with experimental data,
exact transition dipoles calculated by using scattering waves should
be used. This does not pose additional difficulties within the
fitting method. In this connection we note here that this extension
(beyond the plane-wave approximation) would make
the simple tomographic procedure inapplicable, as the Fourier slice
theorem cannot be used directly.

\section{Conclusions and Perspective}

In this paper we have shown that the simple iterative fitting
procedure is very efficient for extracting the internuclear
separations from the high-order harmonics generated by infrared
lasers. This method, in combination with the available quantum
chemistry codes, can be made a very powerful tool for exploring
the time-resolved structural changes in a chemical transformation.
The procedure is much simpler than the tomographic method
suggested by Itatani {\it et al} \cite{itatani}. It also does not
rely on the assumptions made in that paper.

For the iterative method to work, an adequate theory for
calculating   high-order harmonics from molecules has to be
available and the calculations be very efficient. In the present
work, this theory is available in the form of strong field
approximation (SFA). However, the SFA is an approximate theory and
the HHG spectra calculated using SFA is not expected to be in
agreement with the experimental data. In principle, HHG spectra
can be calculated by solving the time-dependent Schr\"odinger
equation. However, such calculations for molecular targets in
arbitrary alignment is very time consuming and the results are
likely not of sufficient accuracy in general. Thus in the near
future, even for the simple molecules such as N$_2$, O$_2$ and
CO$_2$ studied in the present paper, the iterative method based on
Eq.~(3) is not practical. Fortunately, an alternative route is
possible. Very recently it has been shown by Morishita {\it et al}
\cite{toru} and Le {\it et al} \cite{atle08} that Eq.~(1) is
applicable to accurate HHG spectra of atoms generated by infrared
lasers. The establishment of validity of Eq.~(1) in this case was
based on accurate HHG spectra obtained by solving the
time-dependent Schr\"odinger equation for atoms in intense laser
fields and for transition dipole matrix elements calculated using
accurate scattering waves, i.e.,  accurate wavefunctions of
electrons in the continuum. We anticipate that HHG spectra from
molecular targets obtained from accurate theory (for H$_2^+$
target this has been shown in Le {\it et al} \cite{atleH2+})  or
from experiment can also be factored out in the form of Eq.~(1)
where the transition dipole matrix elements $d(\omega,\theta)$ are
calculated using scattering waves. Note that $d(\omega,\theta)$ is
the same transition dipole matrix element calculated in the study
of photoionization cross sections of molecules.

Unlike the
nonlinear laser-molecule interactions, photoionization is a linear
process, and they can be calculated with much less effort. In
particular, for our purpose, we do not need high-precision
photoabsorption cross sections in a narrow energy region such as
those measured with synchrotron radiation, but rather moderately
accurate dipole matrix elements over a broad photon energy range.
Thus simpler calculations based on the one-electron model along
the line developed by Tonzani \cite{tonzani} is likely adequate
for this purpose. Clearly further development along this line
requires the input of experimental data in order to demonstrate
the actual working of the iterative method. If the method is
established, it would be rather straightforward to extend the
method to carry out the time-resolved chemical imaging with
few-cycle infrared laser pulses,  achieving temporal resolutions
down to a few femtoseconds, depending on the pulse durations of
the probe pulses used.

\ack

This work was supported in part by Chemical Sciences, Geosciences
and Biosciences Division, Office of Basic Energy Sciences, Office
of Science, U.S. Department of Energy.


\section*{References}
\begin{thebibliography}{50}
\bibitem{zewail00} Zewail A H 2000 {\it J. Phys. Chem. A} {\bf 104} 5660

\bibitem{zewail01} Ihee H, Lobastov V A, Gomez U M, Goodson B M,
Srinivasan M, Ruan C Y and Zewail A H 2001 {\it Science} {\bf 291} 458

\bibitem{leinreview} Lein M 2007 \JPB {\bf 40} R135

\bibitem{itatani} Itatani J, Levesque J, Zeidler D, Niikura H, Pepen H,
Kieffer J C, Corkum P B and Villeneuve D M 2004 {\it Nature} (London)
{\bf 432} 867

\bibitem{hoang} Le V H, Le A T, Xie R H and Lin C D 2007 \PRA {\bf 76} 013414

\bibitem{moadk} Tong X M, Zhao Z and Lin C D 2002 \PRA {\bf 66} 033402

\bibitem{gaussian} Frisch M J {\it et al} 2003 {\it GAUSSIAN 03, revision C.02}
(Gaussian, Inc., Pittsburgh, PA)

\bibitem{gamess} Schmidt M W {\it et al} 1993 {\it J. Comput. Chem.} {\bf 14} 1347

\bibitem{lewenstein} Lewenstein M, Balcou Ph, Ivanov M Yu, L'Huillier A and
Corkum P B 1994 \PRA {\bf 49} 2117

\bibitem{zhou} Zhou X X, Tong X M, Zhao Z X and Lin C D 2005 \PRA {\bf 71} 061801(R)

\bibitem{atle} Le A T, Tong X M and Lin C D 2006 \PRA {\bf 73} 041402(R)

\bibitem{levesque} Levesque J, Mairesse Y, Dudovich N, Pepin H, Kieffer J C,
Corkum P B and Villeneuve D M 2007 \PRL {\bf 99} 243001

\bibitem{NRC} Pavicic D, Lee K F, Rayner D M, Corkum P B and
Villeneuve D M 2007 \PRL {\bf 98} 243001

\bibitem{litvinyuk} Litvinyuk I V, Lee K F, Dooley P W, Rayner D M,
Villeneuve D M and Corkum P B 2003 \PRL {\bf 90} 233003

\bibitem{cockeN2} Alnaser A S, Voss S, Tong X M, Maharjan C M, Ranitovic P,
Ulrich B, Osipov T, Shan B, Chang Z and Cocke C L 2004 \PRL {\bf 93} 113003

\bibitem{cockeCO2} Alnaser A S, Maharjan C M, Tong X M, Ulrich B, Ranitovic P,
Shan B, Chang Z, Lin C D, Cocke C L and Litvinyuk I V 2005 \PRA {\bf 71} 031403(R)

\bibitem{atle07} Le A T, Tong X M and Lin C D 2007
{\it J. Mod. Optics} {\bf 54} 967

\bibitem{toru} Morishita T, Le A T, Chen Z and Lin C D 2008
\PRL {\bf 100} 013903

\bibitem{atle08} Le A T, Morishita T and Lin C D 2008 submitted
to \PRL

\bibitem{atleH2+} Le A T {\it et al} (in preparation)

\bibitem{tonzani} Tonzani S 2007 {\it Comput. Phys. Comm.} {\bf 176} 146

\end{thebibliography}
\end{document}